\documentclass[prb,twocolumn]{revtex4-1} 

\usepackage{amsmath}  
\usepackage{amsfonts} 
\usepackage{graphicx} 

\begin{document}


\title{Numerical radiative transfer with state-of-the-art iterative
  methods made easy}

\author{Julien Lambert}
\email{julien.lambert@physics.uu.se} 
\affiliation{Lund Observatory, S\"olvegatan 27, Box 43, SE--221 00 Lund, Sweden}
\affiliation{Department of Physics and Astronomy, Uppsala University, Box 516, SE--75120 Uppsala, Sweden}

\author{Fr\'ed\'eric Paletou}
\email{fpaletou@irap.omp.edu}
\affiliation{Universit\'e de Toulouse, OMP, CNRS, IRAP, F--31400 Toulouse, France}

\author{Eric Josselin}
\email{eric.josselin@univ-montp2.fr}
\affiliation{Universit\'e de Montpellier, CNRS, LUPM, F--34095 Montpellier, France}

\author{Jean--Michel Glorian}
\email{Jean-Michel.Glorian@irap.omp.edu}
\affiliation{Universit\'e de Toulouse, OMP, CNRS, IRAP, F--31400 Toulouse, France}


\date{\today}

\begin{abstract}
  This article presents an on-line tool and its accompanying software
  resources for the numerical solution of basic radiation transfer
  out of local thermodynamic equilibrium (LTE). State-of-the-art stationary
  iterative methods such as Accelerated $\Lambda$--Iteration and
  Gauss-Seidel schemes, using a short characteristics-based formal
  solver are used. We also comment on typical numerical experiments
  associated to the basic non--LTE radiation problem. These resources
  are intended for the largest use and benefit, in support to
  more classical radiation transfer lectures usually given at the Master
  level.
\end{abstract}

\maketitle 

\section{Introduction}

The theory of radiation transfer is of paramount importance for
astrophysics. Indeed, except for those objects in the Solar System
close enough to us for being explored \emph{in situ}, from the
collection of lunar samples back in the 70's to the spectacular
landing of the Philae spacecraft and its instruments on-board, on
comet Churyumov--Gerasimenko in November 2014,
our knowledge of the Universe overwhelmingly comes  from the
analysis of the light we collect from \emph{distant} objects.

For the emblematic case of stars, photons are generated in the central
parts of the body. They first scatter across its internal and still
opaque layers, finally escaping the star at the bottom of its
atmosphere or \emph{photosphere}. These photons will continue to be
scattered through the most external layers (and any possible
circumstellar structures therein, like for instance solar prominences
lying in the corona) of the star, before reaching the interstellar
medium and travel into space down to our instruments. Finally, stellar
light will scatter again through the Earth atmosphere, when one
considers ground-based astronomical observations.

The issue of how radiation transfers along these media of distinct
physical nature, in terms of density, temperature, dynamics, magnetic
field etc. thus appears as quite obvious. And even though we shall
discard hereafter any discussion about terrestrial atmospheric
effects, radiation transfer through stellar atmospheres still is a
very difficult problem of physics. It relies indeed on \emph{complex}
non-linear light--matter interactions (see e.g., Hubeny \& Mihalas
2014\cite{mihalas}).

The equation of radiative transfer is very likely to be present in all
astrophysics courses, more likely at the Master level. Analytical
solutions are very few, but they can quite easily be taught and fully
derived within a few lectures of introduction to the radiation
transfer theory. Despite the often very crude approximations
used in these cases, such solutions may still be very useful to any
astronomer willing to ``clutch at straws'' while facing a problem of
interpretation of data (or of numerical results) involving some more or
less complicated radiative modelling.

At first glance, the radiative transfer equation (hereafter RTE)
appears as a deceptively simple first order ordinary differential
equation. This is indeed the case when the so-called \emph{source
  function} is already known, and the process of deriving the
radiation field from a known source function is refered to as the
\emph{formal solution}. However, in the more general case, the RTE is
\emph{integro-differential}, because the source function depends on
integral terms involving the radiation field itself. The general
problem of defining self-consistently the radiation field i.e., the
specific intensity and its first moments, together with the detailed
excitation and ionization states of an atmosphere (and therefore the
\emph{opacity}, as well as the spatial variations of the source
function), considering the highly non-linear light--matter
interactions which usually take place in astrophysical plasmas is
definitely, and still, a (very) \emph{difficult} problem (see e.g.,
Rutily \& Chevallier 2006\cite{rutily}).

A considerable simplification of the problem is brought by the
assumption of considering that the astrophysical plasma permeatted by
radiation is in the physical conditions of the so-called Local
Thermodynamic Equilibrium (hereafter LTE). In such a case, velocity
distributions of particles follow a Maxwell--Boltzmann distribution,
excitation and ionisation stages of every atom (or molecule)
constituting the plasma follow Boltzmann (for excitation equilibrium)
and Saha (for ionisation equilibrium) statistics, and the
self--consistent source functions are, accordingly, characterized
locally by Planck functions (see e.g., Chandrasekhar
1960\cite{chandra}).  More insight about the assumptions underlying
LTE -- and therefore how departures from LTE can take place in a
stellar atmosphere -- can be found in the monographs of Hubeny \&
Mihalas (2014)\cite{mihalas} and Oxenius (1986)\cite{oxenius}.  One
should also realize that \emph{realistic} radiative modelling, even
assuming LTE, is such complex already that it actually requires
numerical modelling.  But even though LTE may be suitable for several
astrophysical ``objects'' (e.g., stellar photospheres),
\emph{departures} from LTE should always be considered as the most
general situation. These effects have indeed been identified and
studied as early as in the late 60's, with the advent of ``numerical
radiation transfer'' (see e.g., Cuny 1967\cite{cuny}, Mihalas \& Auer
1969\cite{cl1,cl2}).  However a few analytical solutions to the
non--LTE radiation transfer problem may be derived and used for the
sake of validating any numerical approach to the problem.

Excluding probabilistic methods such as Monte-Carlo (see e.g., Auer
1968\cite{mc1}, Bernes 1979\cite{bernes} and Whitney 2011\cite{mc} for
a recent review), we may consider that solutions of the non--LTE
radiation transfer equation fall into two main classes being either
difference equation methods (e.g., Mihalas \& Auer 1969), or
\emph{stationary}\cite{NonStat} iterative methods (e.g., Olson,
Auer \& Buchler 1986 and references therein). Hereafter we shall focus
on iterative methods, from the so-called $\Lambda$--iteration (or
Picard) method to the Jacobi-like approximate or \emph{Accelerated}
$\Lambda$--Iteration (also known as ALI) method and, finally, the more
recent and much less popular still Gauss-Seidel and Successive
over-relaxation (SOR) schemes (Trujillo Bueno \& Fabiani Bendicho
1995)\cite{gs}.

The treatment of the RTE thus appears to be also a good introduction
to numerical techniques, including the use of the moments of a
function, in physical sciences.

We first remind basic equations driving the non--LTE (unpolarized)
radiation transfer problem in a static and 1D plane-parallel
geometry. We shall also restrict ourselves to the special case of
monochromatic scattering. Then we shall be able to derive an
analytical solution of the NLTE radiation problem. This solution shall
therefore be used for testing several iterative methods, including the
very popular ALI. Finally, we shall describe a new on-line tool
designed for educational purposes. It is located at {\tt
  http://rttools.irap.omp.eu/}, and the associated Python software
will also be made available from us.

\section{Formal solution of the radiation transfer equation}

The derivation of the RTE in a semi-infinite, plane-parallel, static,
and 1D geometry can be found in several classic textbooks (Hubeny \&
Mihalas 2014\cite{mihalas}, or in the $e$-book of Rutten
2003\cite{rob}).

\begin{figure}[]
  \includegraphics[width=7.25 cm,angle=0]{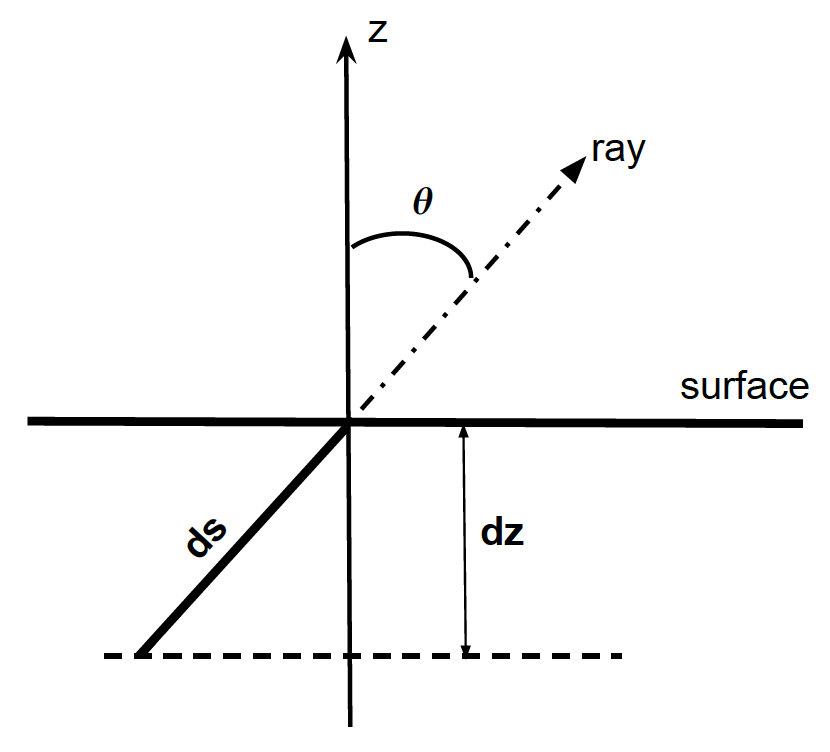}
  \caption{Geometry of the 1D plane-parallel radiative transfer
    problem. A ray emerges from a semi-infinite atmosphere with angle
    $\theta$ vs. the normal to the surface $\vec{z}$. The direction
    cosine is usually called $\mu$. Along the ray, we use the spatial
    coordinate $s$.}
  \label{Fig1}
\end{figure}

The source function is defined as $S_{\nu}=\eta_{\nu}/\chi_{\nu}$
i.e., the ratio bewteen monochromatic emissivity and the extinction or
absorption coefficient.  The optical depth is defined as $d \tau_{\nu}
= - \chi_{\nu} dz$ where $\mu=\cos(\theta)$ and $dz=\mu ds$, assuming
that $\vec{z}$ points in the \emph{opposite} direction of gravity and
that the direction cosine $\mu$ defines the orientation of a ray
vs. the $z$-axis -- see Fig.\,(1). Then RTE comes into its more
familiar form:

  \begin{equation} 
\mu { {dI_{\nu}} \over {d \tau_{\nu}}} = I_{\nu} - S_{\nu}  \, .
  \end{equation} 
where $I_{\nu}$ and $\tau_{\nu}$ also depend on $\mu$.
  This basic RTE is the one which appears in most lectures notes and
  text books introducing radiation transfer to astrophysicists.
    We shall also assume that the source function is isotropic i.e.,
    angle-independent. 

  The so-called ``formal solution" is the general solution of this
  equation for a \emph{known} source function. In such as case, the
  solution can be easily derived as:

\begin{equation}
 I_{\nu}(\tau_1)= I_{\nu}(\tau_2)  e^{-(\tau_2 - \tau_1)/\mu} + 
 \int_{\tau_1}^{\tau_2}   { S_{\nu}(\tau_{\nu}) e^{-(\tau_{\nu} - \tau_1)/\mu}
 \left( d\tau_{\nu}/\mu \right) }
  \label{eq:formal}
\end{equation}

We shall see hereafter how this formal solution is used together with
iterative methods we are particularly interested in. Note also that,
when $\tau_1 \rightarrow 0$ and $\tau_2 \rightarrow +\infty$ in
Eq.\,(\ref{eq:formal}), the formal solution is the \emph{Laplace}
transform of the source function.

\subsection{The Eddington approximation}

It is usual and convenient, for analytical radiation transfer, to
define the three successive (angular) moments of the specific intensity,
or Eddington moments:

\begin{eqnarray}
 J_{\nu} & = & {1 \over 2}  \int_{-1}^{+1} {I_{\nu}(\mu) d \mu }  \\
H_{\nu} & = &{1 \over 2}  \int_{-1}^{+1} {I_{\nu}(\mu) \mu d \mu } \\
K_{\nu} & = & {1 \over 2}  \int_{-1}^{+1} {I_{\nu}(\mu) \mu^2 d \mu }
\end{eqnarray} 
where $J$ is the mean intensity, $H$ the Eddington flux (which relates
to the astrophysical flux usually observed for spatially unresolved
objects such as most stars other than the Sun), and $K$ which is related
to the radiation pressure (see e.g., Hubeny \& Mihalas 2014\cite{mihalas}).

In a similar fashion, successive moments of the RTE can be easily
derived. It leads, respectively for Eddington factors $H_{\nu}$ and $K_{\nu}$ to

\begin{equation}
  { {dH_{\nu}} \over {d \tau_{\nu}}}  = J_{\nu} -S_{\nu}   \, ,
\end{equation} 
and

\begin{equation}
  { {dK_{\nu}} \over {d \tau_{\nu}}}  = H_{\nu}   \, .
\end{equation} 

Then, considering the state of the radiation field \emph{at great depth}
in a stellar atmosphere when it can be safely considered also as
isotropic, one can establish the so-called Eddington approximation,
$J_{\nu}=3K_{\nu}$ (Rutten 2003\cite{rob}, Hubeny \& Mihalas
2014\cite{mihalas}).

Although valid only in the above-mentioned conditions, it is common in
analytical radiation transfer to consider that the Eddington approximation
remains valid throughout the whole atmosphere of a star, even though
the anisotropy of the radiation field increases while we move towards
its most external layers.

\subsection{$\Lambda$-operator}

It is usual in the field of astrophysical radiation transfer to write
the formal solution of RTE as

\begin{equation}
 I_{\nu}(\tau) = \Lambda [ S(\tau) ]
\end{equation}
where the operator $\Lambda$ represents the operation of deriving the
specific intensity from known opacity and source function spatial
distributions. It is also, in other terms, an integration operator of
the known source function weighted by the exponential kernel
$e^{(t-\tau)}$. 

Should we ignore the frequency dependence of the radiation field, also
known as the ``grey case'', and include the angular integration
leading from specific intensity to the mean intensity (i.e., a
physical quantity proportional to the more generally observed
astrophysical flux), one would rather write the formal solution as:

\begin{equation}
 J(\tau) = \Lambda [ S(\tau) ] \, .
\end{equation}

For all computations presented hereafter, as well as for the tools we
distribute, this process is done using the Python module {\tt formal}
which computes the formal solution using the so-called short
characteristics method (Olson \& Kunasz 1987 \cite{sc}, Auer \&
Paletou 1994 \cite{lhafp}).

\section{An analytical solution to a NLTE radiation problem}

An analytical solution to the problem of non--LTE radiative transfer
can be derived with the following assumptions. First, we shall
consider the case of monochromatic or coherent scattering.  We shall
also consider a source function that contains a thermal emission
component $\varepsilon B$ and a coherent isotropic scattering term
$(1-\varepsilon) J$, that is

\begin{equation}
S=  (1-\varepsilon) J+  \varepsilon B \, .
  \label{eq:source}
\end{equation}
In the frame of the two-level atom model, $\varepsilon$ is also called
the collisional destruction probability factor (it may also be related
to a so-called albedo, with $a=1-\varepsilon$ though).

Assuming that the Eddington approximation $J=3K$ is valid \emph{at all
  depths} in the atmosphere, we get easily after forming the second
derivative of $K$ that

\begin{equation}
  { {d^2 J} \over {d \tau^2}}  = 3(J -S)   \, .
\end{equation} 
For an isothermal atmosphere of constant $B$ with $\tau$ and
using the expression of the source function introduced at Eq. (\ref{eq:source}), the
latter expression turns into

\begin{equation}
  { {d^2 {(J-B)}} \over {d \tau^2}}  = 3 \varepsilon (J -B)   \, ,
\end{equation} 
whose solution is such that $(J-B) = \alpha e^{- \tau \sqrt{3
    \varepsilon}}$.

Finally, using the boudary condition at the surface
  $J(0)=\sqrt{3}H(0)$ derived by Krook (1955)\cite{krook}, one can
establish that

\begin{equation}
  \alpha  = - { {B} \over {1 + \sqrt{\varepsilon}} }  \, ,
\end{equation} 
which leads to the Eddington solution of the non--LTE radiation
transfer problem:

\begin{equation}
S_{\rm Edd.}/B= 1 - (1-\sqrt{\varepsilon}) e^{- \tau \sqrt{3 \varepsilon} }  \, .
  \label{eq:sedd}
\end{equation}

It is important to identify two critical values associated with this
solution. First is the surface value, $S_{\rm Edd.}(\tau)$ for $\tau
\rightarrow 0$, of the source function given by $\sqrt{\varepsilon}
B$. Any numerical solution have to be tested vs.  this limit value,
and with some accuracy. Second is the typical depth at which $S_{\rm
  Edd.} \simeq B$ which is often called the \emph{thermalisation
  depth}. It scales as $1/\sqrt{\varepsilon}$ for the monochromatic
scattering case we consider here. Again this typical length should be
identified with accuracy from any numerical solution to the non--LTE
problem.

We conclude this section emphasizing the fact that the Eddington
  solution we have established (for a semi-infinite atmosphere) is the
  \emph{true} solution of an \emph{approximate} radiation transfer
  problem. The main approximation used here is on the angular
  dependence of the specific intensity. While the resolution of the
  full problem would imply to use an infinite number of directions, we
  downsized it to a single point angular quadrature (see also e.g., the
  discussion provided in \S5. of Chevallier et al. 2003). 

\section{Numerical solutions}

We shall consider hereafter monochromatic (or \emph{grey}) radiation
transfer, so we can drop any frequency dependance of the Eddington
factors in the remainder of this article. Boundary conditions are also
assumed to be monochromatic.

The Eddington approximation is also compatible with the so-called
``two-stream approximation''. In that case, we adopt also a simplified
angular quadrature using $\mu=\pm 1/\sqrt{3}$ (i.e., the Van Vleck
angle). Beyond astrophysics, this approximation is also common
for global circulation or weather forecasting models developed in
(terrestrial) atmospheric sciences (\emph{J.-P. Chaboureau}, private
communication).  Note again that a proper comparison between
  numerical solutions and the (analytic) Eddington solution requires
  the use of a single point angular quadrature.

Our formal solver uses short characteristics (SC) using monotonic
parabolic interpolation, as originally described in Auer \& Paletou
(1994 -- see also Olson \& Kunasz 1987\cite{sc}, Kunasz \& Auer
1988\cite{ka88}, and Paletou \& L\'eger
2007\cite{fp07}).\cite{SCtypos}. Hereafter we only remind the
  essential principles of SC, and encourage the reader to consult the
  existing scientific literature, for details. The numerical
  implementation of short-characteristics rely on the following
  principles. Short characteristics means that the formal solution
  across the whole atmosphere will be carried-out \emph{depth after
    depth}, from one boundary surface to the other one, and
  back-and-forth i.e., for $\mu$ negative first then for $\mu$
  positive (note that this order is indifferent, but the separation
  between positive and negative direction cosines is very important,
  and shall prove very useful for the case of Gauss--Seidel
  iterations). In order to perform at each spatial depth the formal
  solution expressed by Eq.\,(2), we shall first assume that the
  source function is \emph{quadratic} in the optical depth. This
  assumption allows to derive an analytical expression of the integral
  term on the source function spatial distribution. Then it can easily
  be shown that, for a current position $k$ the integral in Eq.\,(2)
  will \emph{only} involve quantities known at this very position, and
  at the ``upwind" or $(k-1)$, and ``downwind" or $(k+1)$
  positions. Finally, at each depth the current value of the specific
  intensity (for a given direction cosine) will be given by

\begin{equation}
  I_{k} = I_{k-1}e^{-\Delta \tau_u} + \Psi_u S_u + \Psi_0
  S_0 +\Psi_d S_d  \, ,
  \label{eq:sc}
\end{equation} 
where the $\Psi$'s are analytical functions of the optical depths
$\Delta \tau_u$ and $\Delta \tau_d$ i.e., between the local ($0$), and
the upwind ($u$) and downwind ($d$) spatial positions.  

 For all cases discussed hereafter, we use boundary conditions
  more usually used for ``semi-infinite'' atmospheres. There is
  \emph{no} radiation falling ($\mu<0$) onto the top surface of the
  atmosphere, and the bottom and upward ($\mu>0$) boundary condition
  is that the specific intensity equals the Planck function (set to
  one hereafter).

\subsection{$\Lambda$--iteration}

The so-called $\Lambda$--iteration ({\sc li}) is equivalent to a Picard iterative
scheme (itself being a fixed-point iterative scheme for ODEs).

Let $S^{\dagger}$ be the spatial distribution of the source function
across the atmosphere from (the initial value or) the previous
iteration step. $\Lambda$--iterating consists in successively computing

\begin{equation}
J = \Lambda [ S^{\dagger}]  \, ,
\end{equation}
then $S^{\rm (new)} = (1-\varepsilon) J +  \varepsilon B$ and so on,
until convergence.

Unfortunately, this poor numerical scheme is still in use although it
is well-known that it is ``pseudo-convergent'' (see e.g., Hubeny \&
Mihalas 2014\cite{mihalas}, Olson et al. 1986\cite{oab}). This is well
demonstated by Figs.\,(2) where we displayed (a) the successive
iterates of the {\sc li} scheme together with the target analytical
solution of Eddington and, (b) the respective relative correction,
from an iteration to another, $R_e$ and the so-called ``true error''
$T_e$, which is the relative error vs. the analytical Eddington
solution.

Following the definitions found in the original papers of Auer et
  al. (1994)\cite{mg} and Trujillo Bueno \& Fabiani Bendicho (1995)\cite{gs},
these two latter quantities are respectively

\begin{eqnarray}
 R_{e} & = & {\rm max}(\lvert S^{(n)}-S^{(n-1)} \rvert /S^{(n)} ) \\
T_{e} & = & {\rm max}(\lvert S^{(n)}-S_{\rm Edd.} \rvert /S_{\rm Edd.}
) \, .
\end{eqnarray} 
Both are indeed useful to demonstrate the pseudo-convergent nature and
the \emph{failure} of {\sc li}. Indeed on Fig. (2b) once can notice
the constant drop of the relative correction $R_e$ giving the misleading
impression that pushing the iteration number will finally reach the
solution, while the true error $T_e$ indicates that the
pseudo-solution will remain far away from the reference solution of
Eddington.

\begin{figure}[]
  \includegraphics[width=7.25 cm,angle=0]{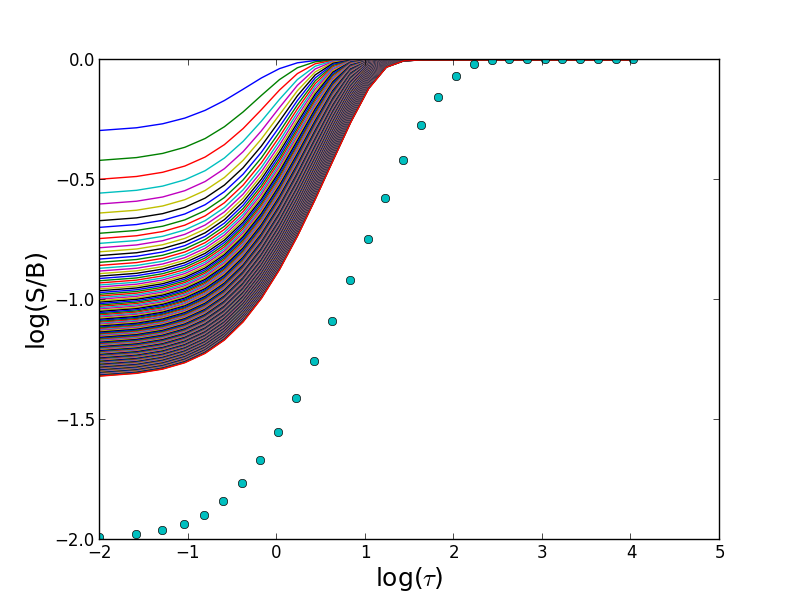}
  \includegraphics[width=7.25 cm,angle=0]{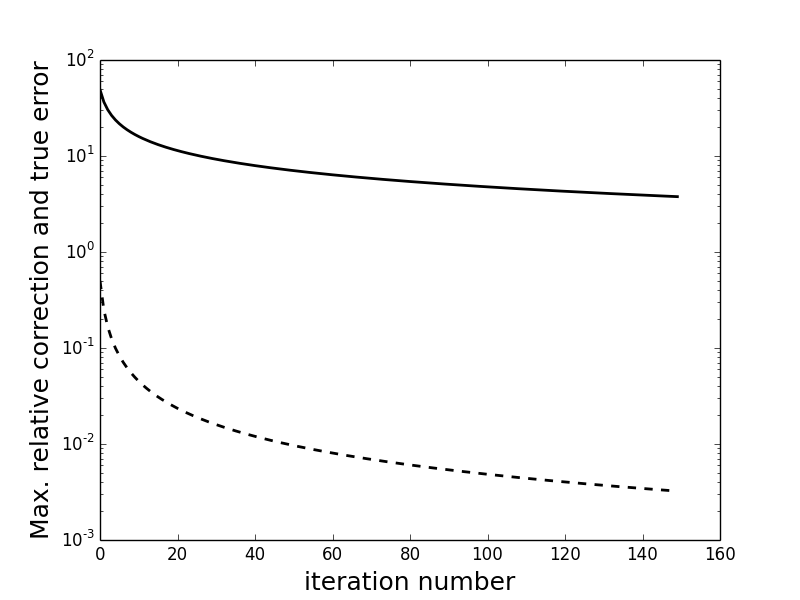}
  \caption{$\Lambda$--iteration for a semi-infinite slab of total
    optical thickness $10^4$ with 5 points per decade, and an
    atmosphere such as $\varepsilon=10^{-4}$. The top figure shows the
    successive runs of the source function, and the pseudo-convergent
    nature of the numerical scheme. The analytical solution is
    represented by the dotted curve. The bottom figure displays (i)
    the maximum relative correction, from an iteration to
    another (dashed line), and (ii)
    the true error i.e., the relative error vs. the analytical
    solution (note that we always have $T_e > R_e$ after a few iterations).}
  \label{Fig2}
\end{figure}

Note also that, in all figures $T_e > R_e$ after the very first
iterative steps, if not at the onset of the iterative process.

\subsection{ALI: Accelerated/Approximate $\Lambda$--iteration}

The {\sc ali} method is basically an operator splitting method. Let us
write therefore:

\begin{equation}
\Lambda = \Lambda^{*}+ \delta \Lambda  \, .
\end{equation}
At this point several choices for an approximate operator
$\Lambda^{*}$ are possible. However, in our study we shall only
consider the most efficient version of {\sc ali} which is just a
Jacobi-type method. In such a case, $\Lambda^{*}$ should be the
\emph{exact} diagonal of the full operator $\Lambda$. The study of
reference concerning this very method is the seminal article of Olson,
Auer \& Buchler (1986)\cite{oab}. In practice, the diagonal operator
is very easily determined, as described in Auer \& Paletou
(1994)\cite{lhafp}

\begin{figure}[]
  \includegraphics[width=7.25 cm,angle=0]{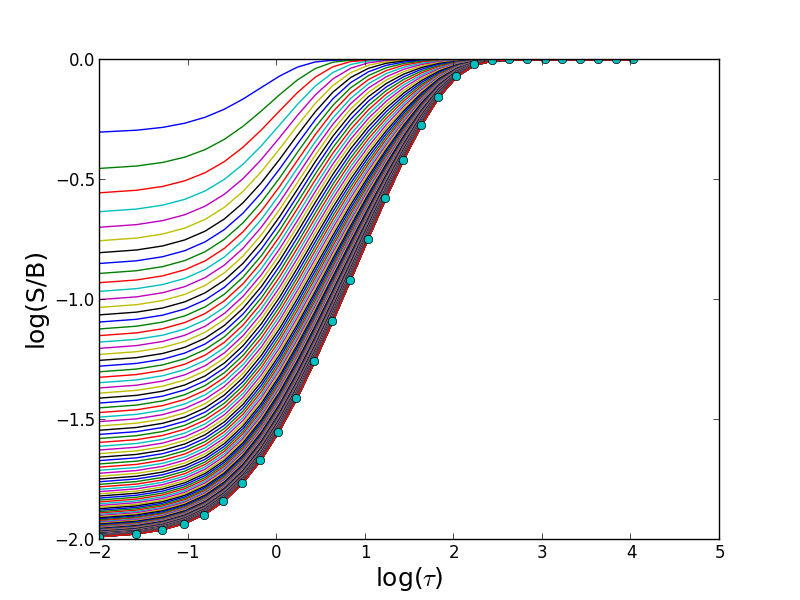}
  \includegraphics[width=7.25 cm,angle=0]{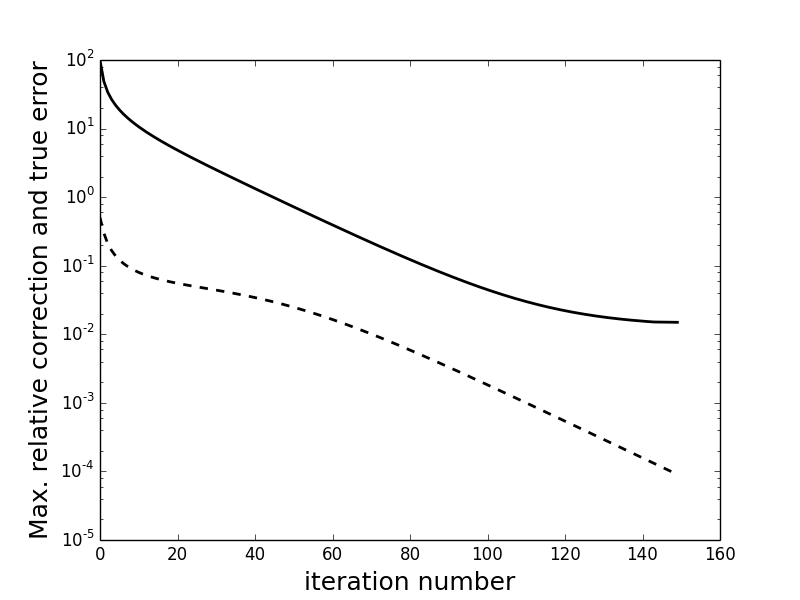}
  \caption{Same as Fig. (2) but for Accelerated
    $\Lambda$--iteration. Within the same number of iterations (150)
    the analytical reference solution is reached, unlike with {\sc
      li}. The true error reaches a constant
    level around 0.01, also indicating that there is no need to
    iterate further than $\sim 140$ iterations, while the relative correction
    continues to decrease.}
  \label{Fig3}
\end{figure}

Now let us write the succession of iterates of the source function as
$S=S^{\dagger} + \delta S$, where $S^{\dagger}$ means the source
function known at the current iterative step. Now using the
  definition of the $\Lambda$-operator $J=\Lambda[S]$ we can
  write the expression of the source function correction explicitely:

\begin{equation}
\delta S = \left[ 1 - (1-\varepsilon) \Lambda^{*}\right]^{-1}  \{
  (1-\varepsilon) \Lambda[S^{\dagger}]+ \varepsilon B - S^{\dagger} \} \, .
\label{eq:dsk}
\end{equation}
When $\Lambda^{*}$ is the diagonal of the full operator, at each depth
in the atmosphere the increment of source function is just obtained
from a \emph{scalar} divide, which make the {\sc ali} method both
accurate and fast.

Unlike the {\sc li} iteration, it can be rigorously shown  that the
{\sc ali} iterative scheme definitely converges to the solution of the
problem, as demonstrated in Fig. (3a). However, the accuracy of the
numerical solution depends on the spatial sampling of the
atmosphere. It can be measured by the true error, $T_e$, which reaches
a \emph{plateau} at $\sim 10^{-2}$, as can be seen in Fig. (3b).

\subsection{Gauss-Seidel and SOR iterations}

Experimenting both Gauss--Seidel ({\sc gs}) and Successive
over-relaxation ({\sc sor}) are logical steps after having
experienced the Jacobi-type {\sc ali} methods. Although published
  twenty years ago now, by Trujillo Bueno \& Fabiani Bendicho
  (1995)\cite{gs}, it is not yet of common practice, unlike {\sc
    ali}.

\begin{figure}[]
  \includegraphics[width=7.25 cm,angle=0]{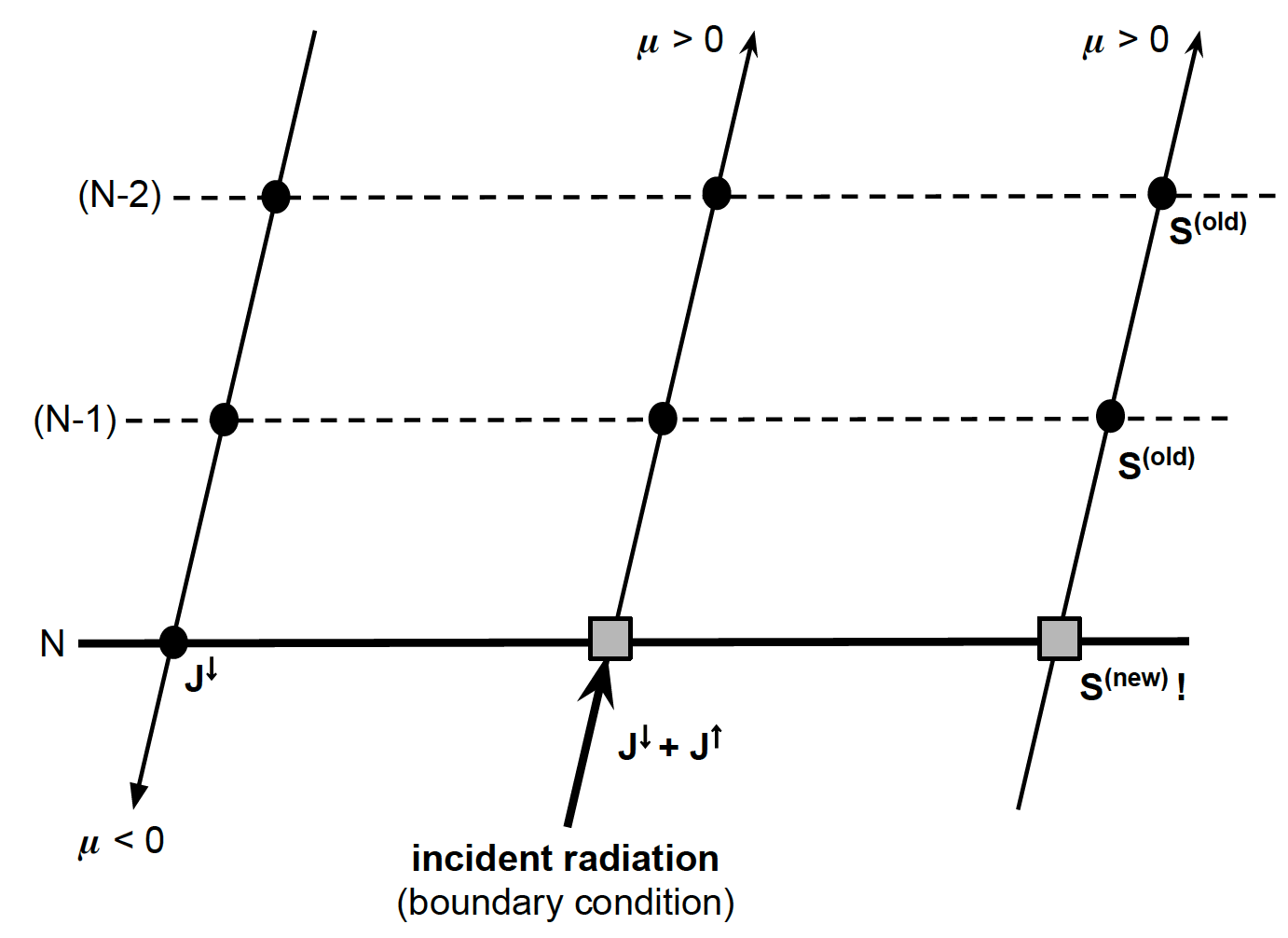}
  \caption{Illustration in support of the basic principle of the
    implementation of Gauss--Seidel iterations within the short
    characteristics method.}
  \label{Fig4}
\end{figure}

 The {\sc gs} iterative method consists essentially in updating
  the current source function value once the full angular integration
  of the specific intensity can be performed -- because all the
  necessary quantities yet are available, and \emph{before} the formal
  solver will be moving to the next depth point. This is made
  relatively easy within the short characteristics methods which
  separates sweping the atmosphere first for $\mu<0$ and second for
  the remainder $\mu > 0$ directions (or vice versa, the important
  point being an explicit distinction between positive and negative
  direction cosines). This is sketched in Fig.\,(4) which should be
  read from left to right. Assume that all depths have been covered
  along SC's of $\mu<0$, so that we know all specific intensities and
  mean intensities $J^{\downarrow}$ for these direction cosines, up to
  the bottom boundary layer $N$. The next task is to complete the
  angular integration for the upward direction cosine(s). Starting at
  the bottom boundary, the ``upwind'' specific intensity is known
  since it is provided by the external boundary condition. Therefore,
  at layer $N$ we can easily compute $J^{\uparrow}$. This knowledge
  makes it possible to update the local source function \emph{before}
  switching to the next inner layer at $(N-1)$. This is the main
  ``trick'' for doing Gauss--Seidel iterations within the SC
  method. It requires however modifications of the more traditional
  formal solver used for ALI (provided by us as the {\tt formalGS}
  module). Indeed, when moving to the next layer for $\mu>0$ at
  $(N-1)$, we shall advance the specific intensity according to
  Eq.\,(\ref{eq:sc}) but now using a mixture of the \emph{just
    updated} $S_u^{\rm (new)}$ and of $S_{0,d}^{\rm (old)}$ i.e., of the
  \emph{not yet} modified values of the source function at the local
  and at the downwind position $(N-2)$. The process is then repeated
  up to the top boundary surface of the atmosphere.

  The numerical implementation of the method was described in every
  detail in the original article of Trujillo Bueno \& Fabiani Bendicho
  (1995)\cite{gs}.  We therefore strongly encourage the reader to
  study this article with great care.


Figure (5) shows the significative gain on the convergence rate
provided by {\sc gs}. For 1D problems, it is by far superior to the
small additional computations induced by the indispensable
modifications of the classical short characteristics formal solver.

\begin{figure}[]
  \includegraphics[width=7.25 cm,angle=0]{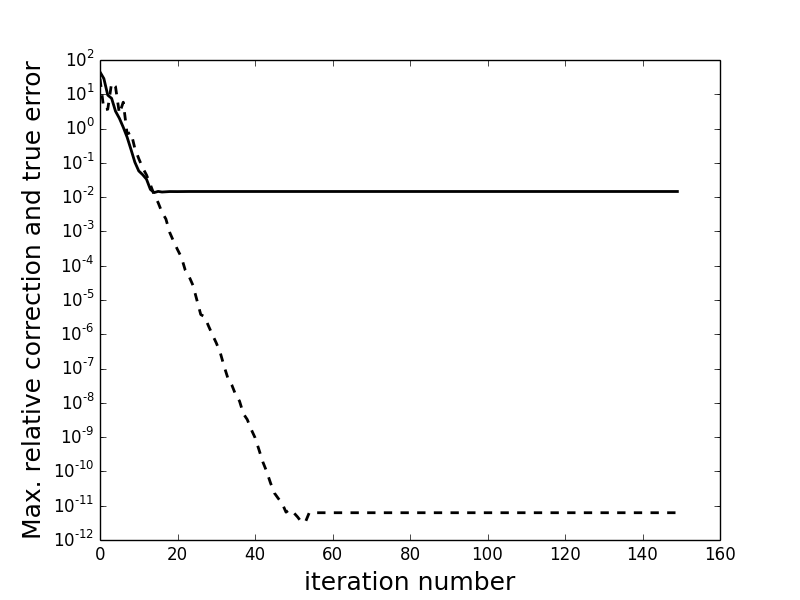}
  \caption{True error and relative correction (dashed
      line) for the same parameters than in previous figures, but
    using the {\sc gs} iterative scheme. The truncation error seen by
    the \emph{plateau} at $T_e \sim 10^{-2}$ is reached in less than
    30 iterations now.}
  \label{Fig5}
\end{figure}

Finally the {\sc sor} scheme is built on the same strategy adopted for
{\sc gs} although the new source function increment $\delta S^{\rm
  (SOR)}=\omega \delta S^{\rm (GS)}$, with $\omega$ chosen between 1
and 2. It can be shown that the optimal scheme is obtained for $\omega
\sim 1.5$ (see Trujillo Bueno \& Fabiani Bendicho 1995\cite{gs}) More
insight about the {\sc sor} method can be found in Young
(1971).\cite{young}

\section{NLTE radiation transfer On-line}

We have implemented a dedicated web-service which allows on-line
numerical experiments with the numerical methods we just presented. It
is located at {\tt http://rttools.irap.omp.eu/}, and maintained by
OMP-IRAP (Toulouse, France). In addition, a Git repository is about to be
  installed, in order to distribute the original Python modules we
  developed and used for the web-service.

\subsection{Description}

Several ``buttons'' may be independently activated. The first choice
is to be made among methods i.e., $\Lambda$--iteration ({\tt LI}),
Accelerated $\Lambda$--iteration ({\tt ALI}), and Gauss--Seidel or SOR
({\tt GS}). Concerning the latter choice, the distinction between both
schemes will be controled by the $\omega$ ({\tt omega})
parameter. Once the method have be selected, the user will have to
provide a value for $\varepsilon$ ({\tt eps}) with format $10^{-p}$
where $p$ is an integer, the total optical thickness of the
atmosphere $\tau_{\rm max}$ ({\tt taumax}), using a format $10^{k}$, and
the number of points per decade ({\tt npdec}) used for setting the
spatial grid for the computations. Finally, the number of iterations
({\tt niter}) to be performed is required.

The output consists in two graphics. The first one provides the
history of the source function (initialized with $S \equiv B$) for the
number of iterations required, with the corresponding Eddington
solution vs. optical depth. The second plot displays both true error and
relative correction (from an iteration to another) vs. the number of iterations.

\subsection{Experiments}

As a preamble to this part, we remind here that comparison with the
analytical solution of Eddington make only sense for the case of
``effectively thick'' slabs, that is such that $\tau_{max} \gg
1/\sqrt{\varepsilon}$, for the case of monochromatic scattering we
only consider hereafter.

A first and \emph{indispensable} experiment is to realize the
\emph{failure} of the $\Lambda$-iteration . The ``pseudo-convergence''
of this method can be noticed by the rapid and continuous fall of the
relative correction $R_e$, while the ``solution'' at which the process is
converging may remain very far from the Eddington solution, as
indicated by the ``true'' error $T_e$. Further experiments pushing the
number of iterations for the same set-up should be done and analysed.

Using the very same set of parameters, the next step is to experiment
the benefit of the diagonal operator {\sc ali} method. The latter
should quickly reach the Eddington solution with good accuracy, unlike
{\sc li}. Other experiments will allow to check the so-called
$\sqrt{\varepsilon}$-law for the surface value of $S/B$, as well as to
identify the ``thermalisation depth'' $1/\sqrt{\varepsilon}$.

However, even {\sc ali} may work at limited accuracy. A relevant
experiment is to iterate the method with a given set of parameters up
to reaching a \emph{plateau} for $T_{e}$. The latter value gives an
indication of the \emph{truncation error} of the method, due to the
spatial discretization of the numerical method. All other parameters
remaining equals, one should then experiment the effects of
\emph{only} modifying the sampling of the slab, by changing the number
of points per decade parameter. Changes in the limit value of $T_{e}$,
together with the rate of convergence should be investigated. Note
that a detailed study on the accuracy of the {\sc ali} method was
published by Chevallier et al. (2003).

Finally, we propose to go beyond the {\sc ali}--Jacobi method with the
Gauss--Seidel ({\sc gs}) and Successive Over-Relaxation ({\sc
  sor}). Its implementation requires several touchy modifications in
the original short characteristics formal solver which require special
attention. We found them pretty well documented in the original
article, and the interested user will get to it by a careful
inspection of the source code that we also deliver with our
web-service. Gauss-Seidel and SOR differ only by the choice of the
relaxation parameter $\omega$. It should be set to $\omega=1$ for
performing {\sc gs} iterations, although it should be picked between 1
and 2 for experimenting {\sc sor} iterations. It is a good exercice to
test various values of $\omega$, seeking for an optimal scheme.

A more insider study, requiring to use directly the Python code we
make available, would be to test the so-called smoothing capability of
GS/SOR methods which plays a crucial role in multi-grid methods (see
e.g., Auer et al. 1994\cite{mg})

\section{Conclusion}

We made available a tool very suitable to any astrophysics Master
programme, or for any astronomer or physicist willing to start an
initiation to state-of-the-art numerical radiation transfer.

It is quite straightforward to upgrade the simple angular quadrature
we used for that study. It would be also possible to implement, from
our Python formal solver, the computation of a more realistic
scattering integral, by integrating an explicitely $\nu$-dependent
mean intensity $J_{\nu}$ weighted by an a priori known absorption
profile (Gaussian/Doppler or Voigt).

Possible evolutions may be, to develop a specific formal solver for
the 1D spherical problem (see e.g., Auer 1984\cite{lha84}), or to
propose a \emph{simple} multi-level atom version following the
so-called Multilevel-ALI method originally developped by Rybicky \&
Hummer (1991, see also Paletou \& L\'eger 2007). Comparisons with
  (non-stationary) conjugate gradient type method could also be
set-up (see e.g., Paletou \& Anterrieu 2009\cite{cg}).

\begin{acknowledgments}

  FP is grateful to his radiative transfer {\it Sensei},
  Dr. L. H. ``Larry'' Auer. We also wish to pay tribute to the
  outstanding contribution of Prof. R. J. ``Rob'' Rutten, whose
  remarkable lectures notes have been made available to anyone for
  more than a decade now. We are also grateful to Dr. J\'er\^ome Ballot
  (IRAP) for a careful reading of a preliminary version of this article.

\end{acknowledgments}

\end{document}